\documentclass[12pt,a4]{article}
\usepackage{graphicx}
\newcommand{\dis}{\displaystyle}
\begin{document}
\begin{flushright} 
CU-TP/00-03
\end{flushright}
\vspace*{1cm}
\begin{center}
{\LARGE Analysis of atmospheric neutrino oscillations in \\
three-flavor 
       neutrinos}\\
\vspace{0.5cm}       
{\large T. Teshima\footnote{E-mail: teshima@isc.chubu.ac.jp}\ \ and T. Sakai\\
{\it Department of Applied Physics,  Chubu University }\\
{\it Kasugai 487-8501, Japan}}
\end{center}
\vspace{0.5cm}
\begin{abstract}
We analyze the atmospheric neutrino experiments of Super-Kamiokande 
(830-920 live days) using the three-flavor neutrino framework with the 
mass hierarchy $m_1\approx m_2\ll m_3$. We study the sub-GeV, multi-GeV 
neutrinos and upward through-going and stopping muons zenith angle 
distributions taking account of the Earth matter effects thoroughly. 
We obtain the allowed regions of mass and mixing parameters 
$\Delta m^2_{23}$, $\theta_{13}$ and $\theta_{23}$. In our present analysis, 
we used the solar neutrino small angle solution and large angle solution for 
$\Delta m^2_{12}$ and $\theta_{12}$. $\Delta m^2_{23}$ is restricted 
to 0.002-0.01\,{eV$^2$} and $\theta_{13}<13^\circ,\ 
35^\circ<\theta_{23}<55^\circ$ in 90\% C.L. For $\theta_{12}$, there is no 
difference between the large angle solution and the small one. From 
$\chi^2$ fit, the minimum $\chi^2=55$ (54 DOF) is obtained 
at $\Delta m^2_{23}=4\times10^{-3}{\rm eV^2}$, $\theta_{13}=10^\circ$ 
and $\theta_{23}=45^\circ$. In a two flavor mixing approximation 
($\theta_{13}=0$), the minimum $\chi^2=61$ (54 DOF) is obtained at $\Delta 
m^2_{23}=3\times10^{-3}{\rm eV^2}$ and $\theta_{23}=45^\circ$. If 
$\theta_{13}=10^\circ$ is real, the detected $\nu_e$ events in K2K 
experiment will be about 10 times as large as events expected in 
$\theta_{13}=0$ case .
\end{abstract}
\vspace{0.5cm}
\hspace*{1cm}PACS number(s): {12.15.Ff, 13.15.+g, 14.60.Pq}
\newpage
\setlength{\baselineskip}{0.33in}
%%%%%%%%%%%%% section 1%%%%%%%%%%%%%%%
\section{INTRODUCTION}
Super-Kamiokande collaboration  has confirmed a neutrino oscillation 
by their atmospheric neutrino experiments \cite{SUPERKAMIOKANDEI}. 
 In their two-flavor mixing analyses of the sub-GeV and the multi-GeV 
zenith angle distribution, it has been obtained that the $\nu_\mu
\leftrightarrow\nu_\tau$ oscillation is preferred to the $\nu_\mu
\leftrightarrow\nu_e$ oscillation and the range of mass parameter $\Delta 
m^2$ is from $10^{-3}{\rm eV}^2$ to $10^{-2}{\rm eV}^2$.
Recently, Super-Kamiokande collaboration has reported that the $\nu_\mu
\leftrightarrow\nu_\tau$ oscillation is maximally $\sin^22\theta=0.9
\mbox{--}1$ and mass parameter $\Delta m^2$ is from $2.5\times10^{-3}
{\rm eV}^2$ to $5\times10^{-3}{\rm eV}^2$, using the sub-GeV, 
multi-GeV neutrino and upward muons zenith angle distribution experiments
(830--920 live days) \cite{SUPERKAMIOKANDEII}. 
\par
However, these results are obtained from the two-flavor neutrino analyses. 
In order to account for the solar neutrino anomaly data together with 
atmospheric neutrino experiments, three flavor neutrinos are necessary 
at least. In three-flavor neutrinos scenario\cite{THREE,TESHIMA} with a 
mass hierarchy $m_1\approx m_2\ll m_3$, there are necessary two mass 
parameters 
$\Delta m^2_{12}$ and $\Delta m^2_{23}$, and three mixing angles 
$\theta_{12}$, $\theta_{13}$ and $\theta_{23}$. Solar neutrino anomaly 
gives constraint on only three parameters $\Delta m^2_{12}$, $\theta_{12}$ 
and $\theta_{13}$. The MSW solution for solar neutrinos predicts the large 
mixing angle solution ($\Delta m^2_{12}=4\times10^{-6}$ \mbox{--} 
$7\times10^{-5}{\rm eV}^2$, $\sin^22\theta_{12}=0.6 \mbox{--} 0.9$) and the 
small mixing angle solution ($\Delta m^2_{12}=3\times10^{-6} $--$1.2
\times 10^{-5}{\rm eV}^2$, $\sin^22\theta_{12}=0.003 \mbox{--} 0.01$) 
for $\theta_{13}=0^\circ\mbox{--}20^\circ$, and these large and small 
mixing angle solutions are merged for $25^\circ\mbox{--}50^\circ$ 
\cite{SOLAR}. The vacuum solution is also obtained as $\Delta m^2_{12} 
\sim 10^{-10}{\rm eV^2}$. However, CHOOZ experiment \cite{CHOOZ} which is 
a terrestrial experiment using reactor neutrinos gives a strong constraint 
$\sin^22\theta_{13}<0.18$ for large 
$\Delta m^2_{23}$ and $\Delta m^2_{23}<0.9\times10^{-3}{\rm eV^2}$ for 
$\sin^22\theta_{13}\sim1$. From the recent many atmospheric neutrino analyses 
\cite{SUPERKAMIOKANDEI,SUPERKAMIOKANDEII,FOGLI,TESHIMA} using either two- 
or three-flavor neutrinos framework, $\Delta m^2_{23}=10^{-3}\mbox{--}
10^{-2}{\rm eV^2}$ is obtained. Therefore, it can be said that the mixing 
angle $\theta_{13}$ is small but not 
necessarily 0. 
\par
In the three-flavor neutrino framework, if the mixing parameter 
$\theta_{13}$ is not zero, it is necessary to consider 
the interplay between two mass parameters $\Delta m^2_{12}$, 
$\Delta m^2_{23}$ and three mixing angles $\theta_{12}$, $\theta_{13}$, 
$\theta_{23}$ for atmospheric neutrino experimental analyses. 
The oscillation term $\dis{\sin^2 1.27\frac{\Delta m^2_{12}}{E}L}$ cannot 
be neglected though $\Delta m^2_{12}\ll\Delta m^2_{23}$,  
because the term is not so small in the sub-GeV experiment 
($E=0.2\sim1.3{\rm GeV}$) of atmospheric neutrino with zenith 
angle $\theta\sim180^\circ$ at which $L\sim10000{\rm km}$. 
In the multi-GeV experiment ($E>1.3{\rm GeV}$) of atmospheric neutrino 
and terrestrial short- and long-baseline experiments, 
the neutrino oscillation term $\dis{\sin^2 1.27\frac{\Delta m^2_{12}}{E}L}$ 
can be neglected. The angle $\theta_{13}$ has been seen to be small, then the 
$\theta_{13}$ is approximated to be $0$ in usual analyses. 
However, in the case considering 
the matter effects of the Earth, above approximation is not appropriate. 
The matter effect is expressed by the induced mass squared 
$A=2\sqrt{2}EG_FN_e=7.59\rho E\times10^{-5}{\rm eV^2}$, where $N_e$ is the 
number density of electrons, $E$ is the energy of neutrinos (measured in GeV) 
and $\rho$ is the matter density (measured in gm/cm$^3$). 
The Earth density $\rho$ is $3.5\sim13$g/cm$^3$ and the multi-GeV 
experiment energy of neutrinos ranges from 1.3GeV to 100GeV. 
Then the value of $A$ go through the value of mass parameter 
$\Delta m^2_{23}$ and the MSW effect occurs in the mixing angle 
$\theta_{13}$ (see section 3). 
\par
In this paper,  we analyze the Super-Kamiokande atmospheric neutrino 
experimental data(850--920 live days) \cite{SUPERKAMIOKANDEII} of sub-GeV 
and multi-GeV neutrino, upward through-going and stopping muons, using the 
three-flavor neutrino framework with the hierarchy $m_1\approx m_2\ll m_3$. 
In previous paper \cite{TESHIMA}, we analyzed the 535 live days data of 
Super-Kamiokande atmospheric neutrino experiment \cite{SUPERKAMIOKANDEI} 
using similar framework to the present analysis, approximating the Earth 
density to be constant. In this present analysis, we will pursue a full 
calculation adopting the varying density of the Earth. 
We will also discuss the long-baseline K2K experiment \cite{LONGBASELINE} 
in the three-flavor neutrino framework. 
%%%%%%%%%%% section 2 %%%%%%%%%%%%%%
\section{NEUTRINO OSCILLATION IN THREE-FLAVOR NEUTRINOS}
\par 
The unitary matrix $U$ which transforms the mass eigenstate neutrinos 
$\nu_\alpha$ to the flavor eigenstate neutrinos $\nu_{l}$ as the formula 
\begin{equation}
\nu_{l}=\sum^3_{\alpha=1} U_{l\alpha}\nu_\alpha, \ \ \ \ l=e,\ \mu, \tau, 
\end{equation}
is parametrized  as follows: 
\begin{eqnarray}
 U&=&\exp{(i\theta_{23}\lambda_7)}\exp{(i\theta_{13}\lambda_5)}
    \exp{(i\theta_{12}\lambda_2)}   \nonumber   \\
  &=&\pmatrix{
      c_{12}c_{13} & s_{12}c_{13} & s_{13} \cr
      -s_{12}c_{23}-c_{12}s_{23}s_{13} & c_{12}c_{23}-s_{12}s_{23}s_{13} & 
      s_{23}c_{13}\cr
      s_{12}s_{23}-c_{12}c_{23}s_{13} & -c_{12}s_{23}-s_{12}c_{23}s_{13} & 
      c_{23}c_{13} \cr
              }, \\
  & & \quad c_{ij}=\cos{\theta}_{ij},\ \  s_{ij}=
 \sin{\theta}_{ij}. \nonumber 
\end{eqnarray}  
in a case disregarding the CP violation. We assume the mass hierarchy 
\begin{equation}
m_1 \approx m_2 \ll m_3, \label{masshie}
\end{equation}
then $\Delta m_{12}^2\ll\Delta m_{13}^2 \simeq \Delta m_{23}^2$. In this mass 
hierarchy, the transition probabilities $P(\nu_{l}\to\nu_{l'})$ can 
be written as
\begin{eqnarray}
P(\nu_{l}\to\nu_{l})&=&1-2(1-2U_{l3}^2-U_{l1}^4-U_{l2}^4+U_{l3}^4)S_{12}-
4U_{l3}^2(1-U_{l3}^2)S_{23}, \\
P(\nu_{l}\to\nu_{l'})&=&P(\nu_{l'}\to\nu_{l})=2(U_{l1}^2U_{l'1}^2+
U_{l2}^2U_{l'2}^2-U_{l3}^2U_{l'3}^2)S_{12}+4U_{l3}^2U_{l'3}^2S_{23},  
\label{trans.prob.}
\end{eqnarray}
where $S_{\alpha\beta}$ is a term representing the neutrino oscillation 
defined as;  
\begin{equation}
S_{\alpha\beta}=\sin^21.27\frac{\Delta m^2_{\alpha\beta}}{E}L.
\end{equation}
Here $\Delta m^2_{\alpha\beta}=|m^2_\alpha-m^2_\beta|$, $E$ and $L$  are 
measured in units 
eV$^2$, GeV and km, respectively. For the mass parameter $\Delta m^2_{12}$ 
and mixing angle $\theta_{12}$, we use values obtained in the solar 
neutrino experimental analyses\cite{SOLAR}: the large mixing angle solution 
$\Delta m^2_{12}=3\times10^{-5}{\rm eV}^2$, $\sin^22\theta_{12}=0.7$, and 
small mixing angle solution $\Delta m^2_{12}=10^{-5}{\rm eV}^2$, 
$\sin^22\theta_{12}=0.005$. It should be noted that the oscillation term 
$\dis{S_{12}=\sin^21.27\frac{\Delta m^2_{12}}{E}L}$ cannot be neglected in 
the sub-GeV experiment of atmospheric neutrino though $\Delta m^2_{12}
\ll \Delta m^2_{23}$, because this oscillation term is not so small in 
sub-GeV neutrino energy ($E=0.2\mbox{---}1.3{\rm GeV}$) with zenith angle 
$\theta\sim180^\circ$($L\sim10000{\rm km}$).
\par
In the atmospheric neutrino experiments, the matters of the Earth have a 
important effect. Matter effect is represented by a term 
$A=2\sqrt{2}EG_FN_e=7.59\rho E\times10^{-5}{\rm eV^2}$ induced by the Earth 
matter, where $\rho$ is the matter density. In the Earth, 
$\rho=3.5\ \mbox{--}\ 13{\rm g/cm^3}$, 
and sub-GeV and multi-GeV neutrino 
energy ranges from 0.1GeV to 100GeV, then the value of $A$ goes through 
those of $\Delta m^2_{12}$ and $\Delta m^2_{23}$ and the resonance 
happens in mixing angles $\theta_{12}$ and $\theta_{13}$ as will be seen in 
Eqs.(8)--(11) and Eqs.(13)--(16). In the sub-GeV neutrino energy, 
$A\sin2\theta_{13}/2\Delta m^2_{23}\ll1$, then we can approximate 
the mixing matrix $U$ as \cite{KUO} 
\begin{equation}
U^M_{\rm sub}=\exp(i\lambda_7\theta_{23})\exp(i\lambda_5\theta_{13})
\exp(i\lambda_2\theta^M_{12}),
\end{equation}
where
\begin{eqnarray}
&&\sin2\theta^M_{12}=\frac{\Delta m^2_{12}}{\Delta m^{M2}_{12}}
                     \sin2\theta_{12}, \\
&&\Delta m^{M2}_{12}=m^{M2}_2-m^{M2}_1,\ \ \Sigma=m^2_1+m^2_2, \\
&&m^{M2}_{1,2}=\frac12\left\{(\Sigma+A\cos^2\theta_{13})\right.\nonumber\\
&&\qquad\qquad\left.\mp\sqrt{(A\cos^2\theta_{13}-\Delta m^2_{12}
      \cos2\theta_{12})^2
      +(\Delta m^2_{12}\sin2\theta_{12})^2}\right\}, \\
&&m^{M2}_3=m^2_3+A\sin^2\theta_{13}.
\end{eqnarray}
On the other hand, $A\sin2\theta_{13}/2\Delta m^2_{23}$ is not so small 
in multi-GeV neutrino and upward muon energy. In this case, the mixing 
matrix can be expanded by powers of a quantity $\Delta m^2_{12}\sin2
\theta_{12}/2m^2_3$ which is very small. In the lowest order, the mixing 
matrix is expressed as  
\begin{equation}
U^M_{\rm multi}=\exp(i\lambda_7\theta_{23})\exp(i\lambda_5\theta^M_{13}),
\end{equation}
where
\begin{eqnarray}
&&\sin2\theta^M_{13}=\frac{\Delta m^2_{13}}{\Delta m^{M2}_{13}}
                     \sin2\theta_{13},\\
&&\Delta m^{M2}_{13}=m^{M2}_3-m^{M2}_1,\ \ \Lambda=\Sigma-\Delta m^2_{12}
    \cos2\theta_{12},\ \ \ \Sigma=m^2_1+m^2_2,\\
&&m^{M2}_{1,3}=\frac12\left\{(m^2_3+\frac{\Lambda}2+A)\right.\nonumber\\
&&\qquad\qquad\left.\mp\sqrt{\left(A-\Delta m^2_{13}\cos2\theta_{13}
       \right)^2+\left(\Delta m^2_{13}\sin2\theta_{13}\right)^2}
       \right\}, \\
&&m^{M2}_2=\frac12(\Sigma+\Delta m^2_{12}\cos2\theta_{12}).
\end{eqnarray}
These expressions correspond to the neutrino case, and  $A$ for the 
anti-neutrino case has an opposite sign of the neutrino case. From these 
expressions, we can recognize 
that a resonance exists at $A\approx \Delta m^2_{13}\cos2\theta_{13}$ 
of $\sin2\theta^M_{13}$. When $\Delta m^2_{13}=3\times10^{-3}{\rm eV^2}$ and 
$\cos2\theta_{13}\sim1$, the resonance occurs at $E\sim{\rm 10\ GeV}$. 
\par
When the density of the Earth matter is constant, the transition probability 
$P^M(\nu_l\to\nu_{l'})$ with the matter effects can be expressed  by 
the expression (\ref{trans.prob.}) with $U$ and $\Delta m^2_{\alpha\beta}$ 
replaced by $U^M$ and $\Delta m^{M2}_{\alpha\beta}$. Actual density of the 
Earth is not constant but the density in shells which compose the Earth 
is almost constant. The net transition probability $P_{\nu_l \to \nu_{l'}}$ 
of neutrinos going through the Earth can be given by connecting the  
transition amplitudes of each shell. The transition amplitude in $k$-th 
shell is expressed as  
\begin{equation}
\dis{T_{l'l}(M_k,x_k)=\sum_\alpha
(U^{M_k})_{l'\alpha}(U^{M_k*})_{l\alpha}\exp{\left(-i\frac{m^{{M_k}2}_\alpha}
{2p}x_k\right)}},
\end{equation}
where $\dis{U^{M_k}}$ represents the mixing matrix containing the $k$-th 
shell matter effects, $m^{M_k}_\alpha$ the mass of $\nu_\alpha$ in the 
$k$-th shell and $x$ the length traveling in the $k$-th shell.
Using this transition amplitude, we get the transition probability as follows:
\begin{eqnarray}
P_{\nu_l \to \nu_{l'}}&=&|\sum_{l_1,l_2,\cdots}T_{l'l_{n-1}}(M_n,x_n)
T_{l_{n-1}l_{n-2}}(M_{n-1},x_{n-1})\cdots\nonumber\\
&&\qquad\qquad\cdots T_{l_{2}l_{1}}(M_2,x_2)T_{l_{1}l}(M_1,x_1)|^2.
\end{eqnarray} 
Calculation using this expression is very complicated  and cannot be carry 
out analytically, then we calculate this numerically.

%%%%%%%%% section 3%%%%%%%%%%%%%%%%%%%%%%%%
\section{NUMERICAL ANALYSES OF ATMOSPHERIC NEUTRINOS}
In this work, we analyze the ratio $N_{\rm Exp}(l)/N_{\rm MC}(l)$ of 
experimentally observed events $N_{\rm Exp}(l)$ and expected events 
$N_{\rm MC}(l)$ without oscillation, where $l$ represents $\mu$-like 
and $e$-like events. 
The zenith angle $\theta$ dependent events ${dN_{\rm Exp}(l)/d\cos\theta}$ 
and ${dN_{\rm MC}(l)/d\cos\theta}$ are defined as
\begin{equation}
{\frac{dN_{\rm Exp}(l)}{d\cos\theta}}=\sum_{\nu_{l'}}\int\epsilon_{l}(E_{l})
       \sigma_{\nu_{l}}(E_{\nu_{l'}},E_{l},\psi)F_{\nu_{l'}}(E_{\nu_{l'}},
       \theta-\psi)
       P^M(\nu_{l'}\to\nu_{l})dE_{\nu_{l'}}dE_ld\cos\psi,
\end{equation}
\begin{equation}
{\frac{dN_{\rm MC}(l)}{d\cos\theta}}=\int\epsilon_{l}(E_{l})
      \sigma_{\nu_{l}}(E_{\nu_l},E_{l},\psi)F_{\nu_{l}}(E_{\nu_{l}},
      \theta-\psi)dE_{\nu_l}dE_{l}d\cos\psi,
\end{equation}
where the summation $\sum_{\nu_{l'}}$ are taken over $\nu_\mu$ and $\nu_e$. 
In these expressions, processes of $\bar{\nu}_\mu$ and $\bar{\nu}_e$ are 
contained. $\epsilon_{l}(E_{l})$ is the detection efficiency of the detector 
for $l$-type charged lepton with energy $E_{l}$. $\sigma_{\nu_{l}}
(E_{\nu_{l'}},E_{l},\psi)$ is the differential cross section of scattering 
$l$ with energy $E_l$ by incident $\nu_{l}$ with energy $E_{\nu_{l'}}=
E_{\nu_{l}}$, where angle $\psi$ is the scattering angle between the 
directions of incident $\nu_{l}$ and scattered $l$. $F_{\nu_{l'}}(E_{\nu_{l'}},
\theta)$ is the incident $\nu_{l'}$ flux with energy $E_{\nu_{l'}}$ produced 
at the atmosphere coming to the detector with zenith angle $\theta$. 
$E_{\nu_l}$ and $\theta$ dependences of $F_{\nu_l}(E_{\nu_l},\theta)$ for 
multi-GeV experiment ($E_{\nu_l}>1.33{\rm GeV}$) are given in 
Refs.~\cite{GAISSER,HONDA}. These dependences including the 
geomagnetic effects for sub-GeV case ($0.2{\rm GeV}<E_{\nu_l}<1.33{\rm GeV}$)
are taken from Ref.~\cite{HONDAP}. Other informations of $\epsilon_l(E_l)$ 
and $\sigma_{\nu_l}(E_{\nu_l},E_l,\psi)$ are given in Ref.~\cite{KAJITA}. 
The upward through-going muons (thru-muons) and stopping muons (stop-muons) 
fluxes are shown in Ref.~\cite{KAJITA}. Typical energies of $\nu_\mu$ 
that produce thru- and stop-muons are 100GeV and 10GeV, respectively.
Explicit calculation of Eqs.~(19) and (20) is  explained precisely in 
Appendix A of the second paper in Ref.~\cite{TESHIMA}.
\par
$P^M(\nu_{l'}\to\nu_{l})$ is the transition probability with the matter 
effects expressed in Eq.~(18). The  Earth consists of 5 shells 
approximately \cite{CARLSON}; the density of the most outside shell 
(${\rm radius}\ r=5991-6371{\rm km}$) is $3.5{\rm g/cm^3}$, 
the next outside shell 
($r=5719-5991{\rm km}$) $4{\rm g/cm^3}$, the middle shell 
($r=3519-5719{\rm km}$) $5{\rm g/cm^3}$, the outer core shell 
($r=1231-3519{\rm km}$) $11{\rm g/cm^3}$, the inner core ($r=0-1231{\rm km}$) 
$13{\rm g/cm^3}$. A number of the shells through which neutrinos pass  
changes with a change of the zenith angle $\theta$. For example, the 
number of the shell is 0 for $0<\theta<90^\circ$ and 9 for 
$\theta=180^\circ$. 

\par
The zenith angle distributions of sub-GeV, multi-GeV neutrino events and upward 
thru- and stop-muons fluxes are given by the SuperKamiokande 850-920 live 
days experiments \cite{SUPERKAMIOKANDEII}. The data are tabulated in Table I. 
These values are taken from the experimental event data and Monte-Carlo 
simulations which are given graphically in Ref.~\cite{SUPERKAMIOKANDEII}.  
$\mu$-like events include the fully contained and partially contained events. 
Errors represent statistical ones only. 

%%%%%%%%%%%%%%%%%%%%%%%%%%%%%%%%%%%%%%%%%%%%%
\begin{table}
\begin{center}
\caption{e-like, $\mu$-like atmospheric neutrino data, and upward muons 
fluxes of SuperKamiokande experiments (850-920 live days data). 
These values are the ratios of experimental data 
and Monte-Carlo simulations which  are obtained from the graphs in Ref.~
\cite{SUPERKAMIOKANDEII}. $\mu$-like events include fully contained and 
partially contained events, and errors represent statistical ones only.}
\vspace*{0.2cm}
\label{table1}
Sub-GeV data\\
\begin{tabular}{ccc}\hline\hline
$\cos\theta$ range & $e$-like event ratio $N_{\rm Exp}/N_{\rm MC}$ & 
$\mu$-like event ratio $N_{\rm Exp}/N_{\rm MC}$ \\ \hline
$-1.0$\mbox{--}$-0.8$  & $1.10\pm0.08$ & $0.66\pm0.05$ \\
$-0.8$\mbox{--}$-0.6$  & $1.10\pm0.08$ & $0.52\pm0.05$ \\
$-0.6$\mbox{--}$-0.4$ & $1.05\pm0.08$ & $0.59\pm0.05$ \\
$-0.4$\mbox{--}$-0.2$ & $0.98\pm0.08$ & $0.67\pm0.05$ \\
$-0.2$\mbox{--}$0.0$  & $0.99\pm0.08$ & $0.64\pm0.05$ \\
$0.0$\mbox{--} $0.2$  & $1.10\pm0.08$ & $0.71\pm0.05$ \\
$0.2$\mbox{--} $0.4$  & $1.09\pm0.08$ & $0.80\pm0.06$ \\
$0.4$\mbox{--} $0.6$  & $0.90\pm0.07$ & $0.82\pm0.06$ \\
$0.6$\mbox{--} $0.8$  & $1.15\pm0.08$ & $0.80\pm0.06$ \\
$0.8$\mbox{--} $1.0$  & $0.94\pm0.07$ & $0.87\pm0.06$ \\ \hline
\end{tabular}
\end{center}
\vspace*{0.2cm}
\begin{center}
Multi-GeV data\\
\begin{tabular}{ccc}\hline\hline
$\cos\theta$ range & $e$-like event ratio $N_{\rm Exp}/N_{\rm MC}$ & 
$\mu$-like event ratio $N_{\rm Exp}/N_{\rm MC}$ \\ \hline
$-1.0 $\mbox{--}$ -0.8$  & $0.83\pm0.18$ & $0.47\pm0.07$ \\
$-0.8 $\mbox{--}$ -0.6$  & $1.19\pm0.18$ & $0.51\pm0.07$ \\
$-0.6 $\mbox{--}$ -0.4$ & $0.97\pm0.15$ & $0.45\pm0.06$ \\
$-0.4 $\mbox{--}$ -0.2$ & $1.18\pm0.16$ & $0.55\pm0.07$ \\
$-0.2 $\mbox{--}$ 0.0$ & $1.04\pm0.14$ & $0.63\pm0.07$ \\
$0.0 $\mbox{--}$ 0.2$  & $0.94\pm0.13$ & $0.81\pm0.08$ \\
$0.2 $\mbox{--}$ 0.4$  & $1.33\pm0.16$ & $0.91\pm0.09$ \\
$0.4 $\mbox{--}$ 0.6$ & $0.85\pm0.15$ & $0.99\pm0.10$ \\
$0.6 $\mbox{--}$ 0.8$ & $1.00\pm0.17$ & $0.85\pm0.09$ \\
$0.8 $\mbox{--}$ 1.0$ & $1.27\pm0.21$ & $1.00\pm0.10$ \\ \hline
\end{tabular}
\end{center}
\vspace*{0.2cm}
\begin{minipage}{8cm}
\begin{center}
Upward through-going $\mu$ data\\
\begin{tabular}{cc}\hline\hline
$\cos\theta$ range & $\mu$ flux ratio $N_{\rm Exp}/N_{\rm MC}$  \\ \hline
$-1.0 $--$ -0.9$ & $0.84\pm0.11$  \\
$-0.9 $--$ -0.8$ & $0.90\pm0.11$  \\
$-0.8 $--$ -0.7$ & $0.72\pm0.09$  \\
$-0.7 $--$ -0.6$ & $0.95\pm0.10$  \\
$-0.6 $--$ -0.5$ & $0.86\pm0.09$  \\
$-0.5 $--$ -0.4$ & $0.76\pm0.08$  \\
$-0.4 $--$ -0.3$ & $0.82\pm0.09$  \\
$-0.3 $--$ -0.2$ & $1.09\pm0.09$  \\
$-0.2 $--$ -0.1$ & $0.96\pm0.08$  \\
$-0.1 $--$  0.0$ & $1.14\pm0.08$  \\ \hline
\end{tabular}
\end{center}
\end{minipage}
\hfill
\begin{minipage}{8cm}
\begin{center}
Upward stopping $\mu$ data\\
\begin{tabular}{cc}\hline\hline
$\cos\theta$ range & $\mu$ flux ratio $N_{\rm Exp}/N_{\rm MC}$  \\ \hline
$-1.0 $--$ -0.8$ & $0.55\pm0.10$  \\
$-0.8 $--$ -0.6$ & $0.50\pm0.09$  \\
$-0.6 $--$ -0.4$ & $0.60\pm0.09$  \\
$-0.4 $--$ -0.2$ & $0.47\pm0.08$  \\
$-0.2 $--$ 0.0$ & $0.75\pm0.10$  \\ \hline
\end{tabular}
\end{center}
\end{minipage}
\end{table}
%%%%%%%%%%%%%%%%%%%%%%%%%%%%%%%%%%%%%%%%%%%%%%%
\par
Since $P^M(\nu_{l'}\to\nu_l)$ is a function of $\Delta m^2_{12}, 
\Delta m^2_{23}, \theta_{12}, \theta_{13}$ and $\theta_{23}$, the ratio 
${(dN_{\rm Exp}(l)}$ ${/d\cos\theta})/({dN_{\rm MC}(l)/d\cos\theta})$ 
of the zenith angle distributions is a function of $\Delta m^2_{12}, 
\Delta m^2_{23}, $ $\theta_{12}, \theta_{13}, \theta_{23}$ and $\theta$. We 
analyze the atmospheric neutrino data fixing the values of parameters 
$\Delta m^2_{12}$ and $\sin^22\theta_{12}$ determined from the solar 
neutrino experiments \cite{SOLAR} as $\Delta m_{12}^2=3\times10^{-5}{\rm 
eV}^2$ and $\sin^22\theta_{12}=0.7$, which corresponds to the large angle 
solution, and $\Delta m_{12}^2=10^{-5}{\rm eV}^2$ and $\sin^22\theta_{12}=
0.005$, which corresponds to the small angle solution. We treat the ratios of 
the zenith angle distributions of the experimental events and the ones of 
Monte-Calro simulation, $(N_{\rm Exp}(l)/N_{\rm MC}(l))_i$, 
where $i$ represents the region number of the bins of zenith angle $\theta$. 
\par
We calculate numerically the  $\chi^2$  defined as
\begin{equation}
\chi^2=\sum_{i,\ l}\frac{\left\{(N_{\rm Exp}(l)/N_{\rm MC}
(l))^{\rm cal}_i-(N_{\rm Exp}(l)/N_{\rm MC}(l))^{\rm data}_i\right\}^2}
{(\sigma_{\rm st})^2_i+(\sigma_{\rm sy})^2_i}.
\end{equation}
For the sub-GeV, multi-GeV and upward thru-muon experiments, the 
summation on $i$ are from 1 to 10 of zenith angle range bins and for the 
upward stop-muon, from 1 to 5. The summation on $l$  are over $\mu$ and $e$ 
for the sub-GeV and multi-GeV experiments. $\sigma_{\rm st}$ represents the 
statistical error and $\sigma_{\rm sy}$ systematic one. We assumed that 
the value of $\sigma_{\rm sy}$ is $5\%$ of an amount of the $N_{\rm MC}$ 
for the sub-GeV and multi-GeV neutrino data, $10\%$ for the upward stop-muon 
data and $20\%$ for the upward thru-muon data. We adopted the rather large 
value of $\sigma_{\rm sy}$ for upward thru-muons, because there exists 
larger uncertainty of neutrino flux for high energy neutrinos.  
We estimated the values of $\chi^2$ 
for the various values of $\Delta m^2_{23}, \theta_{13}$ and $\theta_{23}$. 
\par
\begin{figure}
\begin{center}
   \includegraphics[width=15cm]{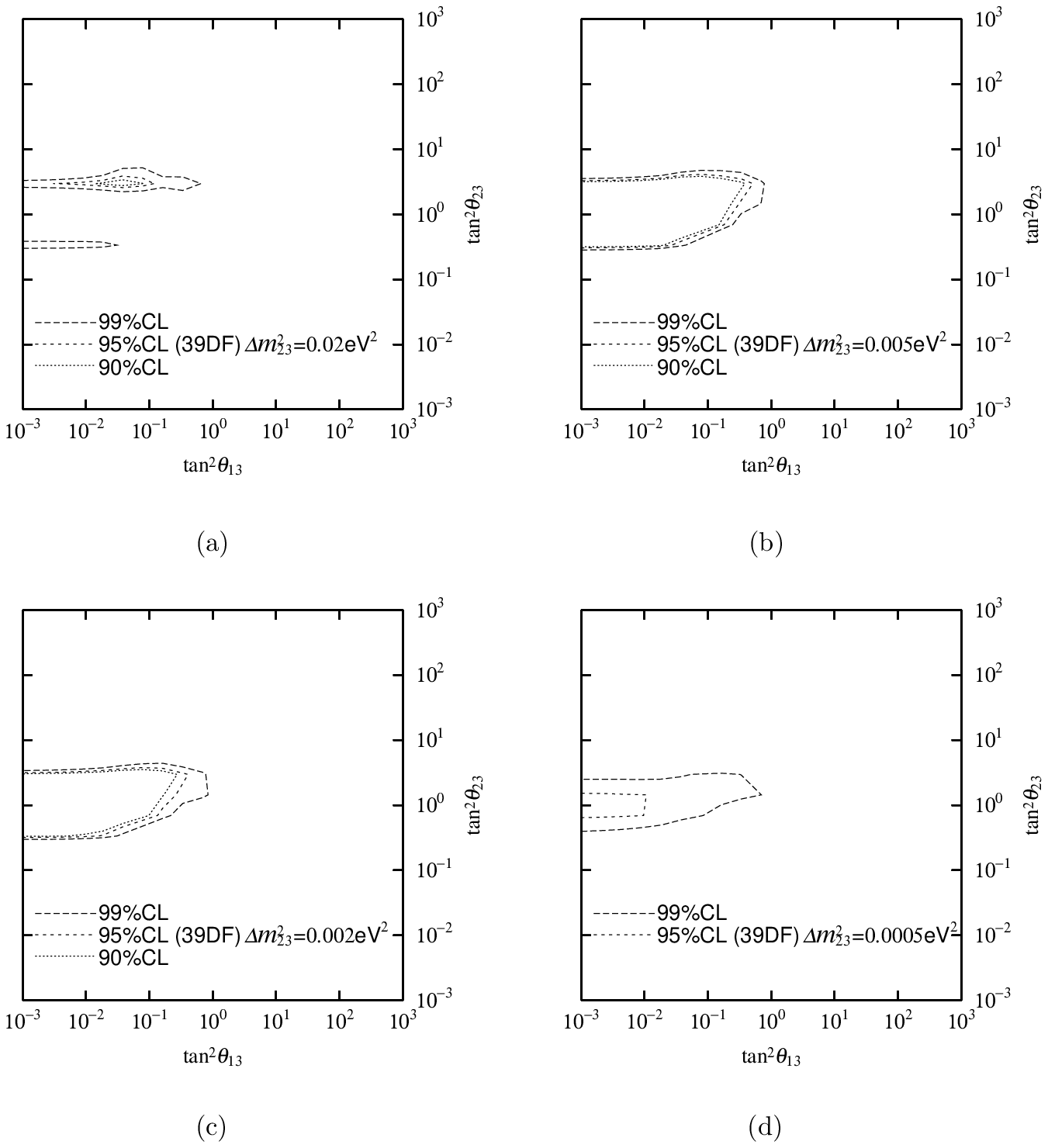}
\end{center}   
\caption{The plots of allowed regions in the $\tan^2\theta_{13}$--$\tan^2
\theta_{23}$ plane determined by the sub-GeV neutrino plus multi-GeV 
neutrino zenith angle distributions of SuperKamiokande 848 live 
days data. These figures correspond to the large 
$\theta_{12}$ angle solution, $\Delta m_{12}^2=3\times10^{-5}{\rm eV}^2$ and 
$\sin^22\theta= 0.7$. In these figures, broken thick, broken thin and dotted 
curves denote the regions allowed in 99\%, 95\% and 90\% C.L., respectively.}
\label{fig1}
\end{figure}

In Fig.~\ref{fig1}, we showed the contour plots of $\chi^2$ 
in the $\tan^2\theta_{13}$--$\tan^2\theta_{23}$ plane for various values of 
$\Delta m^2_{23}$. These plots show the $\chi^2$ of the sub-GeV neutrino plus 
multi-GeV neutrino zenith angle distributions for the large 
$\theta_{12}$ angle solution. We did not show the plots for the small 
angle solution, because  $\chi^2$ for the small angle solution is almost 
similar to that for large one. In these figures, broken thick, broken thin 
and dotted curves denote the regions allowed in 99\%, 95\% and 90\% C.L., 
respectively. From these plots we can say that the values of allowed
$\Delta m^2_{23}$ is  from $2\times10^{-3}{\rm eV^2}$ to $10^{-2}{\rm eV^2}$, 
$\theta_{13}$ is $< 17^\circ$ and $\theta_{23}$ is from $35^\circ$ to 
$55^\circ$.
%%%%%%%%%%%%%%%%%%% fig.1%%%%%%%%%%%%%%%%%%%%%%
%%%%%%%%%%%%%%%%%%%%%%%%%%%%%%%%%%%%%%%%%%%%%%%

%%%%%%%%%%%%%%%%%%% fig.2%%%%%%%%%%%%%%%%%%%%%%
\begin{figure}
\begin{center}
  \includegraphics[width=15cm]{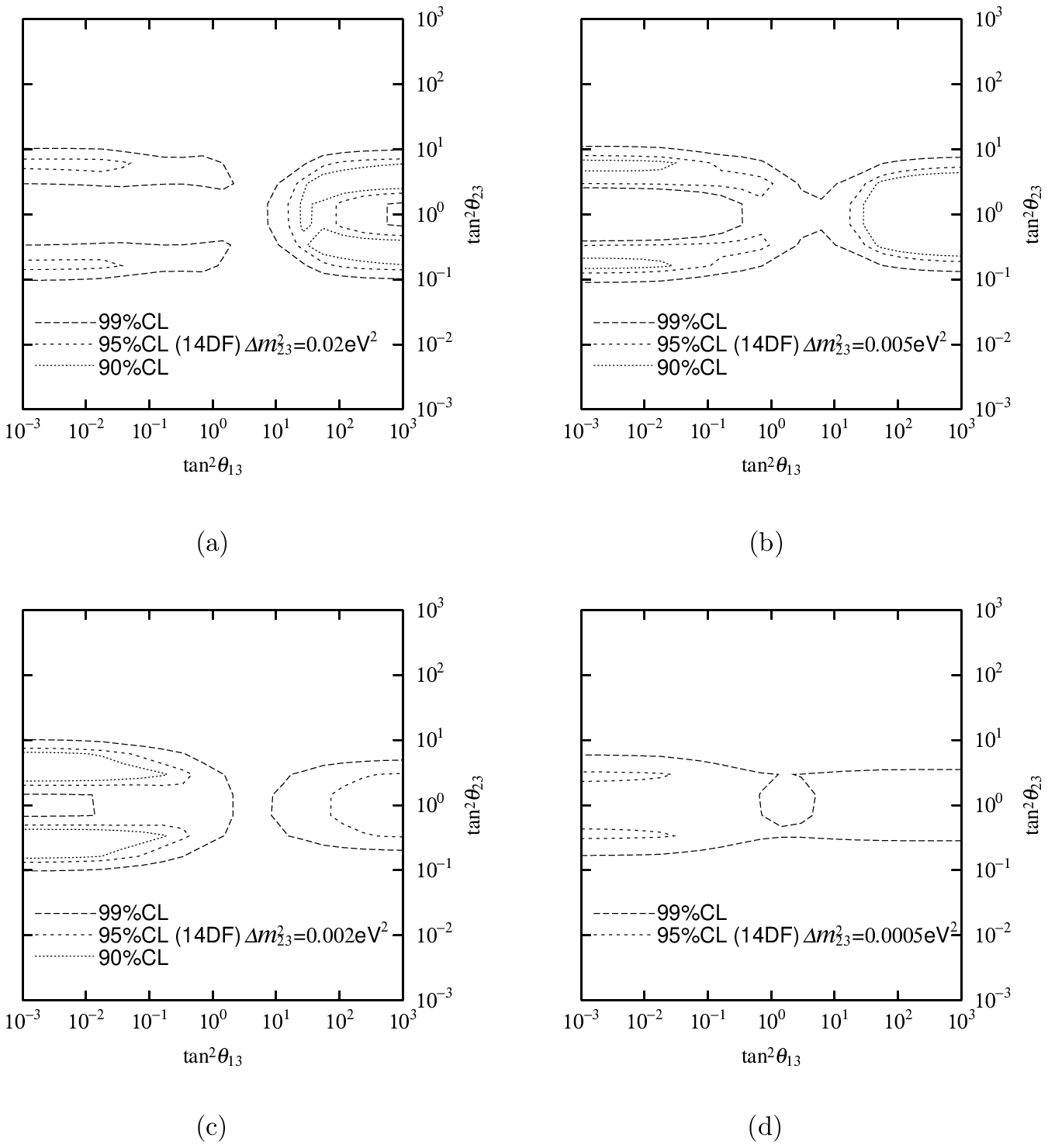}
\end{center}
\caption{The plots of allowed regions in $\tan^2\theta_{13}$--$\tan^2
\theta_{23}$ plane determined by the combination of the zenith angle 
distributions of the upward thru- and stop-muon SuperKamiokande 902-923 
live days data for large $\theta_{12}$ angle solution. There is no 
difference between the large $\theta_{12}$ angle solution and small one. 
In these figures, broken thick, broken thin and dotted curves denote the 
regions allowed in 99\%, 95\% and 90\% C.L., respectively.}
\label{fig2} 
\end{figure}  
%%%%%%%%%%%%%%%%%%%%%%%%%%%%%%%%%%%%%%%%%%%%%%%

%%%%%%%%%%%%%%%%%%% fig.3%%%%%%%%%%%%%%%%%%%%%%
\begin{figure}
\begin{center}
  \includegraphics[width=15cm]{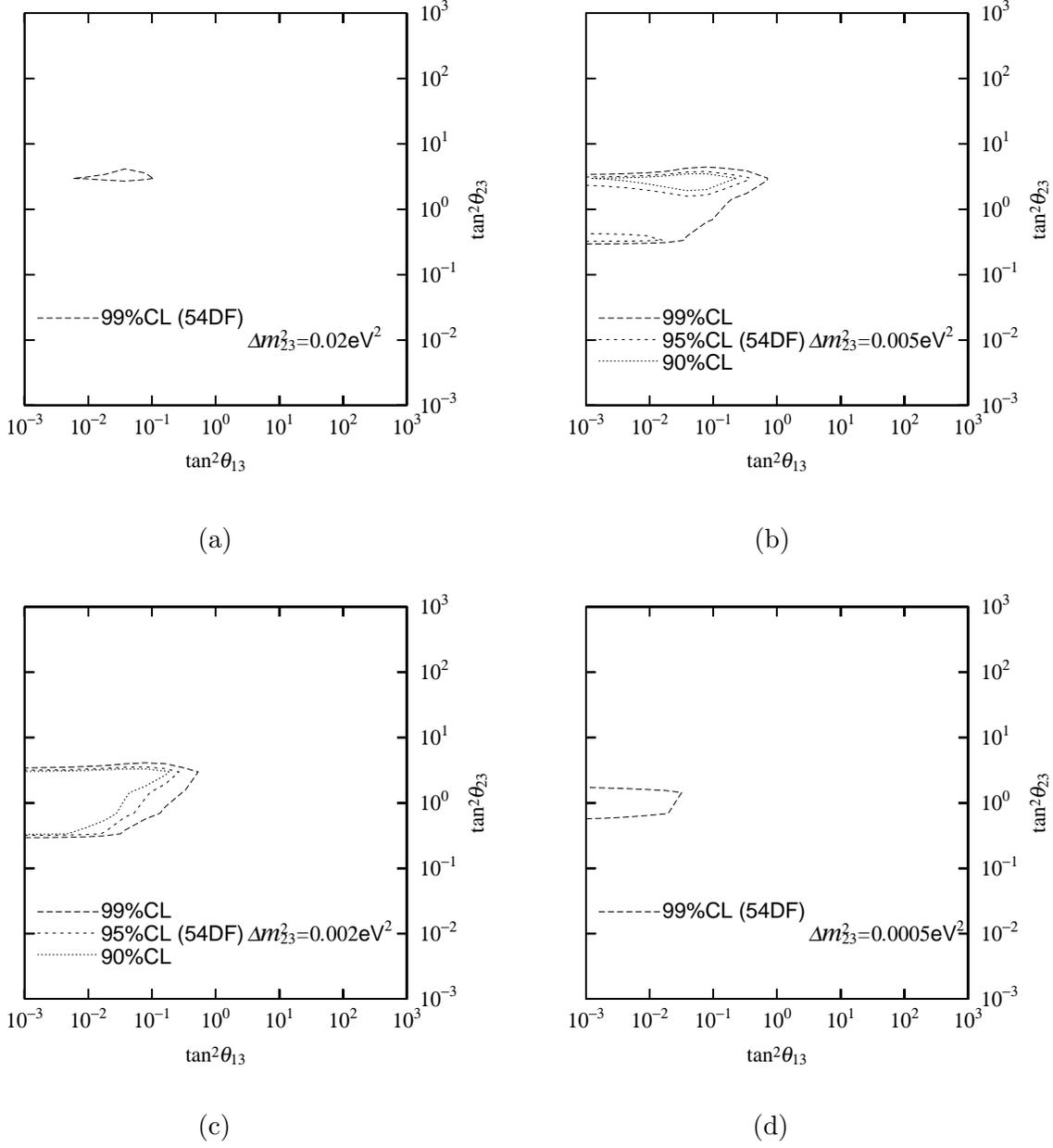}
\end{center}
\caption{The plots of the allowed regions in $\tan^2\theta_{13}$--
$\tan^2\theta_{23}$ plane determined by the combination of the zenith angle 
distributions of the sub- and multi-GeV neutrinos and upward thru- and 
stop-muon data for large $\theta_{12}$ angle solution. There is no 
difference between the large $\theta_{12}$ angle solution and small one. 
In these figures, the broken thick, broken thin and 
dotted curves denote the regions allowed  in 99\%, 95\% and 90\% C.L., 
respectively. }
\label{fig3}
\end{figure}  
%%%%%%%%%%%%%%%%%%%%%%%%%%%%%%%%%%%%%%%%%%%%%%%
\par
Plots of the upward thru- plus stop-muons are shown in Fig.~\ref{fig2} 
for the large $\theta_{12}$ angle solution. We did not show the plots for 
small angle solution, because there is no difference between the 
large $\theta_{12}$ angle solution and small one. In these figures, sharp 
allowed regions similar to those of previous sub-GeV plus multi-GeV 
neutrinos data is not appeared. The reason is as follows: although stop-muons 
data show the sharp allowed regions similar to those of sub-GeV plus 
multi-GeV neutrino data, thru-muons data show the excluded regions near 
$45^\circ$ of $\theta_{23}$. The latter situation is caused from  
small deficit (about 0.2) from 1 of the $\mu$ 
event ratio $N_{\rm Exp}/N_{\rm MC}$ as shown in Table 1. 
\par 
In Fig. 3,  we showed the contour plots of $\chi^2$ for the combination of 
the sub-GeV and multi-GeV neutrinos, the upward thru- and stop-muons zenith 
angle distributions for large $\theta_{12}$ solution. In these figures, 
broken thick, broken thin and dotted curves denote the regions allowed 
in 99\%, 95\% and 90\% C.L., respectively.
\par
From these plots, we can get the following results for the neutrino mass 
and mixing parameters:\\
\ \ (1) As shown in Fig.~1, the allowed region for $\Delta m_{23}^2$ 
obtained from the sub-GeV and multi-GeV neutrino experiments is 
from $2\times10^{-3}{\rm eV^2}$ to $10^{-2}{\rm eV^2}$ at 90\%C.L., 
and the allowed region for $\theta_{23}$ angle is from $35^\circ$ 
to $55^\circ$ and for $\theta_{13}$ angle is less than $17^\circ$. 
The minimum $\chi^2$ is obtained as $29$ for 39 degrees of freedom (DOF) at 
$\Delta m^2_{23}=5\times 10^{-3}{\rm eV^2}$, $\theta_{13}=10^\circ$ and 
$\theta_{23}=45^\circ$. The minimum $\chi^2$ in the restriction 
$\theta_{13}=0$ which corresponds to the two flavor mixing 
$\nu_\mu-\nu_\tau$ approximation, is $30$ for 39 DOF at $\Delta m^2_{23}=
4\times 10^{-3}{\rm eV^2}$. \\
\ \ (2) As shown in Fig.~2, the allowed region for $\theta_{13}$ and 
$\theta_{23}$ is excluded at small $\theta_{13}$ and the maximal mixing of 
$\theta_{23}$ for large values of $\Delta m^2_{23}$. This is because 
of the fact that the data in upward thru-muons has not so large deficit 
from the no-oscillation case as seen in the Table 1.\\
\ \ (3) There is no significant difference between the large $\theta_{12}$ 
solution and the small solution. This fact is not seen from these figures, 
but we can see  from our numerical calculation that the difference between 
the large angle solution and the small one is less than 1\% for multi-GeV 
neutrinos and upward muons and less than 10 \% for sub-GeV neutrinos.\\
\ \ (4) The allowed region for $\Delta m^2_{23}$ obtained from the 
sub- and multi-GeV neutrinos, and upward thru- and stop-muons is given 
as $2\times10^{-3}{\rm eV^2}$ to $\times10^{-2}{\rm eV^2}$, and for 
$\theta_{23}$ as $35^\circ$ to $55^\circ$ and for $\theta_{13}$ as less than 
$13^\circ$. The minimum $\chi^2$ 
is obtained as $55$ for 54 DOF at $\Delta m^2_{23}=4\times 10^{-3}{\rm eV^2}$, 
$\theta_{13}=10^\circ$  and $\theta_{23}=45^\circ$.  The minimum $\chi^2$ 
in the restriction  $\theta_{13}= 0^\circ$ which corresponds to the two 
flavor mixing $\nu_\mu-\nu_\tau$ approximation, is $61$ for 54 DOF at 
$\Delta m^2_{23}=3\times 10^{-3}{\rm eV^2}$ and $\theta_{23}=45^\circ$. 
These results are the same as the result obtained by Super-Kamiokande 
collaboration \cite{SUPERKAMIOKANDEII}. 
\\
\ \ (5) It is interesting that the minimum of $\chi^2$ is obtained at not 
$\theta_{13}=0$ but $\theta_{13}=10^\circ$, though the difference between 
$\chi^2$ for $\theta_{13}=0$ case and that for $\theta_{13}=10^\circ$ case is 
not so large. This result is consistent with the CHOOZ experiment 
\cite{CHOOZ}. The CHOOZE experiment predicts the results: $\theta_{13}<
13^\circ$ for $\Delta m^2_{23}=3\times10^{-3}{\rm eV^2}$. 
\par 
We will now discuss about the interesting feature that the mixing angle 
$\theta_{13}$ is preferred to be about $10^\circ$. If this feature is real, 
the detected $\nu_e$ events in the long baseline K2K experiment 
\cite{LONGBASELINE} will be about 10 times as large as the events 
expected in $\theta_{13}=0$ case. This is caused from the fact that 
the $\nu_\mu$ flux produced at KEK is about 100 times as large as 
$\nu_e$ flux and the transition probability is $P(\nu_\mu\to\nu_e)
\sim\sin^22\theta_{13}\sin^21.27\Delta m^2_{23}L/E$, 
$E\sim1.4{\rm GeV}$ and $L=250{\rm km}$.
%%%%%%%%%%%%%% section 4 %%%%%%%%%%%%%%%%%%%%%%%%
\section{Conclusion}
We analyzed the atmospheric neutrino experimental data of Super-Kamiokande 
\cite{SUPERKAMIOKANDEI} in the three-flavor neutrino framework with the 
mass hierarchy $m_1\approx m_2\ll m_3$ and obtained the allowed regions of 
parameters $\Delta m_{23}^2,\ \theta_{13}$ and 
$\theta_{23}$, including the Earth matter effects thoroughly. 
We studied the event ratios of the sub- and multi-GeV, and 
upward thru- and stop-muons zenith angle distributions. From these 
atmospheric experiments, we can get the allowed region of mass parameter 
$\Delta m^2_{23}$ restricted as $0.002{\rm eV^2}\mbox{--}0.01{\rm eV^2}$, 
and mixing parameter $\theta_{13}$ as less than $13^\circ$ and $\theta_{23}$ 
as $35^\circ\mbox{--}55^\circ$. The value of $\Delta m^2_{23}$ at the 
minimum $\chi^2=55$ for 54 DOF is obtained as $4\times10^{-3}{\rm eV^2}$ at 
$\theta_{13}=10^\circ$ and $\theta_{23}=45^\circ$. 
The minimum $\chi^2=61$ for 54 DOF is obtained with the restriction 
$\theta_{13}=0$ at the $\Delta m^2_{23}=3\times10^{-3}{\rm eV^2}$ and 
$\theta_{23}=45^\circ$. This fact seems very interesting for us because 
the mixing parameter $\theta_{13}$ may not be 0. For mixing parameter 
$\theta_{12}$, the difference between the large angle solution and the 
small one is less than 1\%. If $\theta_{13}$ is about $10^\circ$, the 
detected $\nu_e$ events in K2K experiment is about 10 times as large as 
events expected in $\theta_{13}=0$ case.

%%%%%%%% references %%%%%%%%%%%%%%%%%%%%%%%%%

\end{document}